\documentclass[a4paper]{jpconf}
\usepackage{amsfonts}
\usepackage{amssymb}
\usepackage{amsmath}
\usepackage{bm}
\usepackage{xcolor}
\usepackage{epsfig}
\usepackage{ifthen}
\usepackage{graphicx}
\usepackage{multirow}
\usepackage{float}

\def\noin{\noindent}

\def\be{\begin{equation}}
\def\ee{\end{equation}}
\def\bea{\begin{eqnarray}}
\def\eea{\end{eqnarray}}
\def\bse{\begin{subequations}}
\def\ese{\end{subequations}}
\def\ba{\begin{array}}
\def\ea{\end{array}}

\def\i{\text{i}}

\def\ee{ \end{equation} }

\def\to{\rightarrow}

\def\small{}

\def\g              {\gamma}

\begin{document}
\title{The dispersion method and dimensional regularization applied to the decay \boldmath $H \to Z \gamma$}

\author{I. Boradjiev}
\address{Institute of Solid State Physics, Bulgarian Academy of Sciences, Tzarigradsko chauss\'{e}e 72, 1784 Sofia, Bulgaria}
\author{E. Christova}
\address{Institute for Nuclear Research and Nuclear Energy, Bulgarian Academy of Sciences, Tzarigradsko chauss\'{e}e 72, 1784 Sofia, Bulgaria}
\author{H. Eberl\footnote{speaker}}
\address{Institute for High Energy Physics, Austrian Academy of Sciences, Vienna, Austria}

\ead{helmut.eberl@oeaw.ac.at}

\begin{abstract}
We have calculated the $W$-loop contribution to the amplitude of the decay $H \to Z \gamma$ in the unitary gauge through the dispersion method
and in the $R_\xi$ gauge using dimensional regularization (DimReg). We show that the results of the 
calculations with DimReg and the dispersion method, adopting the boundary condition at the limit $M_W \to 0$
defined by the Goldstone boson equivalence theorem (GBET), completely coincide. This implies that 
the dispersion method obeying the GBET is compatible with DimReg. 
The advantage of the applied dispersion method is that we work with finite quantities and no regularization is required.
\end{abstract}

\section{Introduction}

The $W$-loop contribution to the Standard Model (SM) decay rate of $H \to \gamma \gamma$ has become a subject of a controversy.
This decay is loop induced. Thus it is ultraviolet (UV) finite. 
The individual $W$ amplitudes are UV divergent. Therfore, most authors use dimensional regularisation (DimReg) for the calculations of the loop integrals.
Direct computation within the unitary gauge with 4-dimensional loop integrals by manipulating the sum of the integrands in order 
to have it UV finite, is also possible. 
These two approaches lead to different results.

The process $H \to \gamma \gamma$ is automatically included in $H \to  Z \gamma$ as a limit.
Working with the dispersion integral approach no regularisation is necessary. 
The SM  has a broken SU(2) $\otimes$ U(1) symmetry. As a consequence, the massive vector bosons $Z$ and $W^\pm$ 
have three polarisations. We will see that the contraversial results are directly connected with this feature.

In this contribution based on \cite{Boradjiev18} we will consider the decay $H \to Z \gamma$ and focus on the 
$W$-loop contribution. We first will calculate the $W$ amplitude with the dispersion method and
then as a check with the commonly used DimReg in general $R_\xi$~gauge. The dispersion method is performed in two steps.
First the imaginary part of the amplitude is derived within the unitary gauge and secondly 
the real part of the  amplitude is calculated with the help of the dispersion integral.

\section{The amplitude in the unitary gauge}

\begin{figure*}[htb]\centering
\includegraphics[width=0.9\columnwidth]{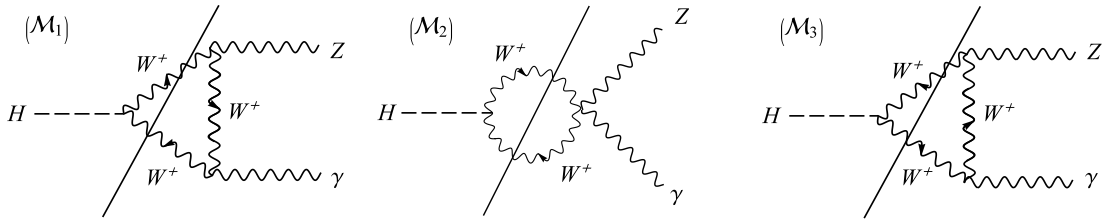}
\caption{Feynman vertex diagrams for the $W$-loop contribution to the decay
$H\rightarrow Z + \gamma$. The inclined lines indicate the cuts.}
\label{Feynman_Diagrams}
\end{figure*}

In the unitary gauge there are three vertex graphs which contribute to the $W$~bosons loop-induced
amplitude of the decay $H\rightarrow Z + \gamma$. They are depicted in Fig.~\ref{Feynman_Diagrams}.
The unitary cuts, needed for obtaining the absorptive parts of the amplitude are shown.
There are two additional diagrams that contribute to $H\to Z + \g$. These are
$H\to Z+Z^*$ with the subsequent transition $Z^* \to \gamma$ with
$W^+ W^-$ and $W^+$ in the loops. Clearly, kinematically their
contribution to the absorptive part is zero and we don't consider
them further.

\noin
The $W$-matrix element for $H(p) \to Z(k_1) + \g(k_2)$ has the form
\begin{align}
{\cal M}=\mathcal{M}_{\mu\nu}(k_1,k_2)\,\epsilon^{*\mu}_1\,\epsilon^{*\nu}_2 \, ,
\end{align}
with
\begin{align}
\mathcal{M}_{1\mu\nu} & = \frac{-\i eg^2\cos\theta_WM}{(2\pi)^4}\int \text{d}^4k \,
\frac{V_{\mu\rho\beta}(-k_1, -P_2, P_1) \,
V_{\nu\gamma\sigma}(-k_2,-P_3, P_2)}{D_1 D_2 D_3} \times \notag \\
&\hspace*{3cm} \times \left(g_{\alpha}^{\beta} - \frac{P_{1\alpha}P_1^{\beta}}{M^2}\right)
\left(g^{\rho\sigma} - \frac{P_2^{\rho}P_2^{\sigma}}{M^2}\right)
\left(g^{\alpha\gamma} - \frac{P_3^{\alpha}P_3^{\gamma}}{M^2}\right) \, ,\\
%\label{def_M1}  
\mathcal{M}_{2\mu\nu} &= \frac{\i eg^2\cos\theta_WM}{(2\pi)^4}\int \text{d}^4k
\frac{V_{\gamma\beta\mu\nu}}{D_1 D_3}
\left(g_{\alpha}^{\beta} - \frac{P_{1\alpha}P_1^{\beta}}{M^2}\right)
\left(g^{\alpha\gamma} - \frac{P_3^{\alpha}P_3^{\gamma}}{M^2}\right)  \, ,
%\label{def_M2}
\end{align}
and 
$\mathcal{M}_{3\mu\nu}(k \to -k)  = \mathcal{M}_{1\mu\nu}$.
The $W$ mass is denoted by $M$, the 3- and 4-vector particle coupling structures $V$ can be found in
Appendix~A of~\cite{Boradjiev18}, 
\begin{align}
P_1 &= k + \frac{p}{2} \, , \qquad P_2 = k - \frac{v}{2} \, , \qquad  P_3 = k - \frac{p}{2}\, ,
\end{align}
with $p =k_1+k_2$, $v=k_1-k_2$, 
$- g^{\mu\nu} + {P_i^\mu P_i^\nu \over M^2}\, (i = 1,2,3)$ are the physical polarisation sums
of the internal $W$~bosons, $D_i = P_i^2 - M^2 + i\epsilon$.
We can write $\mathcal{M}_{\mu\nu}$ as
\begin{equation}
\mathcal{M}_{\mu\nu} = 2 \mathcal{M}_{1\mu\nu} + \mathcal{M}_{2\mu\nu} \, .
\end{equation}

\section{Absorptive part of the amplitude}

Using Cutkosky rules the momenta of the $W$'s are set to be on-shell:
\begin{align}
\frac{1}{p^2 - M^2 + i \epsilon} \longrightarrow \,
(2\pi i) \, \theta(\pm p_0) \, \delta(p^2 - M^2)\, .
\end{align}
\noin
The absorptive part $\mathcal{A}$ is defined by the imaginary part of  $ \mathcal{M}_{\mu\nu}$,
\begin{align}
\Im m \, \mathcal{M}_{\mu\nu} = \frac{eg^2\cos\theta_W}{8\pi M}
\, \mathcal{A}(\tau) \mathcal{P}_{\mu\nu} \, ,
\end{align}
with the transverse factor ${\cal P}_{\mu\nu}=k_{2\mu}k_{1\nu}-(k_1\cdot k_2) \, g_{\mu\nu}$,
and
\begin{align}
\mathcal{A}(\tau) \mathcal{P}_{\mu\nu}
&= \frac{M^2}{\pi} \int \text{d}^4k \,
\mathcal{I}_{\mu\nu} \, \theta(P_{10})\theta(-P_{30})\delta(D_1)\delta(D_3) \, ,
\end{align}
with
\begin{align}
\mathcal{I}_{\mu\nu}&=
\frac{8M_Z^2}{M^4D_2} \, k^2\left(k_{\mu}k_{\nu} + \frac{k_{2\mu}k_{\nu}}{2} -
\frac{k_{\mu}k_{1\nu}}{2} - \frac{k_{2\mu} k_{1\nu}}{4}\right) + \frac{-2M_Z^2}{M^4} \,
k^2g_{\mu\nu} \notag \\
&+ \frac{8M_Z^2}{M^2D_2} \, \left[-k_{\mu}k_{\nu} - \frac{k_{2\mu}k_{\nu}}{2} +
\frac{k_{\mu}k_{1\nu}}{2} - \frac{k_{2\mu} k_{1\nu}}{8} +
\frac{1}{4} \, g_{\mu\nu}k_1\cdot k_2 - \frac{1}{8} \, g_{\mu\nu}k\cdot (k_1-k_2)\right]   + \frac{M_Z^2}{M^2} \, g_{\mu\nu} \notag \\
&+ \frac{2}{M^2D_2} \left[4 k_1\cdot k_2 \, k_{\mu}k_{\nu} + 2k^2k_{2\mu}k_{1\nu}
- 4k\cdot k_1 \, k_{2\mu}k_{\nu} - 4 k\cdot k_2 \, k_{\mu}k_{1\nu}\right. \notag \\
&\left.\hspace*{7cm}+ g_{\mu\nu}\left(4k\cdot k_1 \, k\cdot k_2 - 2k^2 \,
k_1\cdot k_2\right) \right]  \notag \\
&+ \frac{2}{D_2} \bigg[ \left(-3k^2 + 3k\cdot k_1 - 3k\cdot k_2 -\frac{9}{2} \,
k_1\cdot k_2 + 3M^2 - \frac{3}{4} \, M_Z^2\right)g_{\mu\nu} \notag \\
&\hspace*{7cm} + 12k_{\mu}k_{\nu} + 3k_{1\nu}k_{2\mu} - 6k_{\mu}k_{1\nu} +
6k_{2\mu}k_{\nu} \bigg] \, ,\label{Integrand_I}
\end{align}
$\tau = \frac{p^2}{4M^2}$, $a = \frac{M_Z^2}{4M^2}$, and $p$ is the momentum of Higgs boson. 
Note that just in the final result we will set $p^2 = m_H^2$.\\
Using the integrals given in Appendix~B of~\cite{Boradjiev18} we get the non-zero result
\begin{align}
&\mathcal{A}(\tau) = \,  \frac{a}{\tau-a} \left\{
\left[ 1 + \frac{1}{\tau-a} \left( \frac{3}{2} - 2a\tau \right) \right] \beta  \right. \notag \\
& \hspace{2cm}\left. - \left[1 - \frac{1}{2(\tau-a)} \left(2a-\frac{3}{2\tau}\right) - \frac{3}{2a}
\left(1-\frac{1}{2\tau}\right)\right] \ln \left(\frac{1+\beta}{1-\beta}\right) \right\} \, ,  \quad\tau > 1\,  ,
\end{align}
with $\beta = \sqrt{1-\tau^{-1}}$.

\section{The full amplitude}

We define the full amplitude $\mathcal{F}(\tau ,a)$ by
\begin{align}
\mathcal{M}_{\mu\nu} = - \frac{e g^2\cos\theta_W}{8\pi M} \,
\mathcal{F}(\tau ,a) \mathcal{P}_{\mu \nu}  \, .
\end{align}
\noin
The {\it unsubtracted} amplitude ${\cal F}_{un}(\tau ,a)$ 
we calculate from the absorptive part $\mathcal{A}$ by using the 
(convergent) dispersion integral
\begin{equation}
\mathcal{F}_{un}(\tau, a) = \frac{1}{\pi} \int_{1}^{\infty}\frac{\mathcal{A}(y)}{y-\tau}
\, {\rm d}y\, , \quad \tau <1 \, . 
\label{DI}
\end{equation}
\noin
However, $\mathcal{F}_{un}(\tau ,a)$ defines the full amplitude $\mathcal{F}(\tau ,a)$ up to an additive constant $C(a)$:
\begin{equation}
2\pi\,\mathcal{F}(\tau ,a)= 2\pi\,\mathcal{F}_{un}(\tau ,a) +{\cal C}(a)\, .
\end{equation}

\noin
${\cal C}(a)$ still has to be fixed by an appropriate physics condition.
 
 \noin
Using the integrals given in Appendix C of [1] we get the result for $\mathcal{F}_{un}(\tau ,a)$,
\begin{align} 
 & 2\pi\mathcal{F}_{un}(\tau ,a) = \,\, \frac{3 - 4a^2}{\tau-a}  \, + \left( 6 - 4a - \frac{3 - 4a^2}{\tau-a} \right) F(\tau, a)
- 2a \left( 2 + \frac{3 - 4a\tau}{\tau-a} \right) G(\tau, a)\, ,
\end{align}
$F$ and $G$ denote loop integrals and can be found in [1].\\

\noindent
$\mathcal{F}_{un}(\tau ,a)$ has the properties:
\begin{itemize}
\item it is finite at the threshold $\tau = a$
\item it vanishes for $\tau \to \infty$ with fixed $a$
\item for $a \to 0$ we get the corresponding amplitude for $H \to \gamma \gamma$
\end{itemize}

\section{\boldmath $C(a)$ from GBET}
We determine the subtraction constant ${\cal C}(a)$ through the charged ghost contribution adopting the 
{\bf G}oldstone {\bf B}oson {\bf E}quivalence {\bf T}heorem~\cite{GBET} 
which implies that at $M_W \to 0 $, i.e. at $\tau \to \infty$,
the SU(2) $\otimes$ U(1) symmetry of the SM is restored
and the longitudinal components of the physical $W^\pm$~bosons
are replaced by the physical Goldstone bosons $\phi^\pm$.
In the following $\mathcal{M}^\phi_{\mu\nu}$ denotes the amplitude of $H\to Z + \gamma$ in which the
$W^\pm$ are replaced by their Goldstone bosons $\phi^\pm$. The GBET implies
\begin{equation}
\lim_{\tau \to \infty} \,\mathcal{M}_{\mu\nu} (\tau ,a)= \lim_{\tau \to \infty} \,
\mathcal{M}^\phi_{\mu\nu}(\tau ,a)\label{GBET}\,.
\end{equation}
\noin
We calculate  the charged ghost contribution in two different ways:
through direct calculations and via the dispersion integral.
Both calculations lead to the same result.\\

\noin
The Feynman graphs we get from Fig.~\ref{Feynman_Diagrams}
by substituting all internal $W^+$ lines by Higgs ghost $\phi^+$ lines. 
Again, by applying Cutkosky rules to the amplitude we get 
\begin{align}
\Im m\,{\cal M}_{\mu\nu}^{\phi}(\tau ,a)=-\frac{e.g.^2\cos\theta_W}{8\pi
M}\, \frac{M_H^2}{4M^2} \,{\cal A}^\phi (\tau ,a) \,{\cal P}_{\mu\nu} \, ,
\end{align}
with 
\begin{align}
{\cal A}^\phi (\tau ,a)=(1-2a)\,
\frac{2 a\beta -\ln\frac{1+\beta}{1-\beta}}{2\,(\tau -a)^2}\, .
\end{align}
The dispersion integral is
\begin{align}
& {\cal F}^\phi (\tau ,a) = \frac{1}{\pi}\,\int_1^\infty \frac{{\cal A}^\phi (y,a)}{y-\tau } \text{d}y \, ,
\label{disint_phi}
\end{align}
and we obtain
\begin{equation}
\lim_{\tau \to \infty}{\cal F}^\phi (\tau ,a)=  \frac{2\,(1-2a)}{2\,\pi\,\tau} \, .
\end{equation}
From Eq.~(\ref{GBET}) we can deduce 
\begin{align}
\lim_{\tau \rightarrow \infty} 2\pi {\cal F} (\tau,a) = \lim_{\tau
\rightarrow \infty} 2\pi\, [\tau {\cal F}^\phi (\tau,a)] = {\cal C}(a) \, .
\end{align}
Thus we determine $C(a) = 2(1 - a)$.\\

\noindent
An important note: 
In principle, in Eq.~(\ref{disint_phi}) there can be also an additive constant $C^\phi$. But the 
$H \phi^+ \phi^-$ coupling is proportional to $m_H^2$ and not to $\tau$. Consequently the 
large-$\tau$ behavior of ${\cal F}^\phi(\tau ,a)$ goes as ${\cal O}(\tau^{-x})$ with $x \geq 1$.
Therefore the curve integral
$(1/\pi)\int_{ARC}\text{d}y {\cal F}^\phi (y ,a)/(y-\tau)$ over the
infinite arc in the complex $\tau$-plane is zero, $C^\phi = 0$. Furthermore, 
there is no physics reason as GBET for $ {\cal F}(\tau ,a)$.

\section{Calculation in \boldmath $R_\xi$ gauge}
In the $R_\xi$ gauge we have 24 individual vertex graphs and 10 with $Z - \gamma$ selfenergy transition.
The calculation is done automatically with the help of the Mathematica Packages {\tt FeynArts}~\cite{Hahn_01} and {\tt FormCalc}~\cite{Hahn_99}
in DimReg. The $\xi$-independence of the total $W$~amplitude is checked. 
The result ${\cal F}_{\rm DimReg} (\tau ,a)$ coincides with the "classical" one~\cite{BH}.
In the limit $a \to 0$ we also can deduce the result for $H \to \gamma \gamma$~\cite{EGN}.\\

\noin
We get the relation
\begin{align}
2\pi\mathcal{F}_{\text{DimReg}}(\tau ,a) =  2\pi\mathcal{F}_{un}(\tau ,a) + 2(1-2a)\, .
\label{differs}
\end{align}
It is seen that both calculations agree, obeying the GBET.

\section{The decay width}

Approximating the total width by top and $W$-boson loop we get~\cite{BH}
\begin{align}
&\Gamma(H\rightarrow \, Z + \gamma) = \frac{M_H^3}{32\pi} \,
\left(1 - \frac{M_Z^2}{M_H^2}\right)^3 \,  \left[\frac{eg^2}{(4\pi)^2M}\right]^2  
\left| - \cos\theta_W[2\pi\mathcal{F}_W(\tau)] +
c_t \, [2\pi\mathcal{F}_t(\tau_t)]\right|^2 \, ,
\end{align}
with $c_t = \frac{2 \left(3-8\sin^2\theta_W\right)}{3\cos\theta_W}$, 
$\mathcal{F}_t(\tau_t)$ stands for the sum of the $t$ quark one-loop diagrams, and
$\mathcal{F}_W(\tau)$ stands for the sum of the $W$ boson one-loop diagrams.\\

Inserting numerical values we get 
$\Gamma(H \rightarrow \, Z + \gamma) = 6.6 \,\text{keV}$ using $\mathcal{F}_W(\tau) = \mathcal{F}_{\rm un}(\tau,a)$, and 
$\Gamma(H \rightarrow \, Z + \gamma) = 8.1 \,\text{keV}$ using $\mathcal{F}_W(\tau) = \mathcal{F}(\tau,a) = \mathcal{F}_{\rm DimReg}(\tau,a)$.
This means neglecting $C(a)$ the result is smaller by 20\% compared to the correct one. For the decay $H  \to \g \g$ we 
even get a 52\% reduction neglecting $C(0)$.\\

\section{Some references}

The decay width of $H \to \g \g$ was calculated the first time in 1976 in~\cite{EGN} in DimReg and is also 
discussed in \cite{SVVZ}.
In~\cite{HS} the dispersion approach was applied to this channel. More 
recently \cite{MV} discuss how to get the correct result by comparing different techniques.
Furthermore it is shown in \cite{marciano11} and also then in \cite{dedes2012} how to get 
the correct result even within the unitary gauge, and \cite{J} comments
on the important role of the decoupling theorem.\\

\noindent
The decay width of the crossed channel $Z \to H \g$ was calculated in \cite{CCF} and in 1985  in~\cite{BH}
the decay width of $H \to Z \g$ was calculated the first time, using DimReg.\\

\noindent
In \cite{GWW,GWW15,CT} $\Gamma(H \to \g \g)$ again was calculated and in  
\cite{Wu_2017} also $\Gamma(H \to Z \g)$. The results of these four works do not obey GBET,
because they fix $C(a) = C(0) = 0$.\\

Before we conclude it is interesting to discuss the integral, which is the root of the difference giving the "classical" result 
or the one where $C(a)$ is neglected. It has the form
\begin{equation}
I_{\mu\nu} = \int {{\rm d}^n k \over (2 \pi)^n} {4 k_\mu k_\nu - k^2 g_{\mu\nu} \over (k^2 - M_W^2)^3}\, .
\end{equation}
Staying in 4 dimensions, $n = 4$, we get $I_{\mu\nu} = 0$. Calculating the integral in $D$~dimensions, $n = D$, and
performing the limit back to 4 dimensions we then get  $I_{\mu\nu} = {i \over 2} \pi^2 g_{\mu\nu}$. 
It is known that the result has to obey GBET. This is fulfilled when the loop integrals are evaluated by using 
DimReg.

\section{Conclusions}
The $W$-boson induced one-loop contributions to the decay width of $H \to Z + \gamma$ in the Standard Model
have been calculated in the unitary gauge by using the dispersion method.
The result for the decay width of $H \to \gamma \gamma$ is automatically included.\\[-3mm]

\noin
The {\it plus} of our approach is, we deal only with finite quantities which does not involve any uncertainties related to regularization,
and working in the unitary gauge we effectively deal with only 4 Feynman diagrams,
while in the $R_\xi$-gauge one has to calculate 34 (24 for $H \to \gamma \gamma$) graphs.\\[-3mm]

\noin
The {\it minus} is, the amplitude is determined merely up to an additive subtraction constant.\\[-3mm]

\noin
This subtraction constant we have fixed by using the Goldstone Boson Equivalence Theorem.\\[-3mm]

\noin
As a cross-check we also have calculated the amplitude in the commonly used $R_\xi$-gauge class with
dimensional regularization as regularization scheme. We have got the same results as in our dispersion method. \\[-3mm]

\noin
Neglecting the subtraction constant we numerically get for $\Gamma(H \to Z  \gamma)$ a 20\% smaller result 
and for $\Gamma(H \to \g \g)$ even a 52\% smaller result.

\section*{Acknowledgement}
E. Ch. is supported by the Grant 08-17/2016 of the Bulgarian Science Foundation.

\end{document}